# Ultra-broadband mid-infrared emission from Pr$^{3+}$/Dy$^{3+}$ co-doped selenide-chalcogenide glass fiber spectrally shaped by varying the pumping arrangement

**LUKASZ SOJKA,**[2] **ZHUOQI TANG,**[1] **DINUKA JAYASURIYA,**[1] **MEILI SHEN,**[1] **DAVID FURNISS,**[1] **EMMA BARNEY,**[1] **TREVOR M. BENSON,**[1] **ANGELA B. SEDDON,**[1] **SLAWOMIR SUJECKI,**[1,2*]

[1]*Mid-Infrared Photonics Group, George Green Institute for Electromagnetics Research, Faculty of Engineering, University of Nottingham, University Park, Nottingham NG7 2RD, UK*
[2]*Department of Telecommunications and Teleinformatics, Faculty of Electronics, Wroclaw University of Science and Technology, Wybrzeze Wyspianskiego 27, 50-370 Wroclaw, Poland*
*\*Corresponding author: slawomir.sujecki@nottingham.ac.uk*

**Abstract:** In this contribution, a comprehensive experimental study of photoluminescence from Pr$^{3+}$/Dy$^{3+}$ co-doped selenide-chalcogenide multimode fiber samples is discussed. The selenide-chalcogenide multimode fiber samples co-doped with 500 ppm of Pr$^{3+}$ ions and 500 ppm of Dy$^{3+}$ ions are prepared using conventional melt-quenching. The main objective of the study is the analysis of the pumping wavelength selection on the shape of the output spectrum. For this purpose, the Pr$^{3+}$/Dy$^{3+}$ co-doped selenide-chalcogenide multimode fiber samples are illuminated at one end using pump lasers operating at the wavelengths of 1.32 μm, 1.511 μm and 1.7 μm. The results obtained show that the Pr$^{3+}$/Dy$^{3+}$ ion co-doped selenide-chalcogenide multimode fiber emits photoluminescence spanning from 2 μm to 6 μm. Also it is demonstrated that, by varying the output power and wavelength of the pump sources, the spectral shape of the emitted luminescence can be modified to either reduce or enhance the contribution of radiation within a particular wavelength band. The presented results confirm that Pr$^{3+}$/Dy$^{3+}$ co-doped selenide-chalcogenide multimode fiber is a good candidate for the realization of broadband spontaneous emission fiber sources with shaped output spectrum for the mid-infrared wavelength region.

## 1. Introduction

Broadband mid-infrared (MIR) sources are very important for sensing, imaging and communication [1-3]. One of the most promising materials for developing such sources are low phonon energy, rare earth ion doped glasses. Consequently, rare earth ion doped selenide-chalcogenide glasses are the subject of considerable research interest. These materials have a transparent window from ~1 μm to ~20 μm and have vibrational cut-off energies smaller than 350 cm$^{-1}$ [4]. Rare earth ions doped into chalcogenide glasses exhibit large and broad absorption and emission cross-sections and long radiative lifetimes. In addition, it was demonstrated that such materials can be drawn into optical fibers [4]. These features make chalcogenide glasses promising materials for mid-infrared applications [4-10]. Many devices for MIR applications based on chalcogenide glass fibers have been already demonstrated; these include supercontinuum sources and Raman lasers [3, 5]. Moreover, extensive effort has been invested into the design and modeling of MIR fiber lasers based on Pr$^{3+}$, Dy$^{3+}$ and Tb$^{3+}$ ion doped chalcogenide glass fibers [6-9]. However, MIR laser action has not yet been achieved, which is most probably due to the relatively high level of loss present in active chalcogenide glass fibers [10]. Nevertheless, in 2015, an undoped Ge-As-Se-Ga fiber with a minimum loss of 1.6 dB/m at 4.6 µm wavelength was presented by Tang *et al.* [11].

Recently, rare earth ion doped mid-infrared chalcogenide fibers were applied to develop spontaneous emission sources for sensing applications [12-15]. Such sources have low complexity and are cost effective. A spontaneous emission source consists of an active fiber and a pump laser. The pump light excites lanthanide ions to higher energy levels. The excited ions return to lower energy levels whilst emitting photons spontaneously. Thus, the mid-infrared light is produced *via* the spontaneous emission phenomenon and population inversion is not required. The fiber is typically pumped at one end whilst the emitted MIR light is collected from the other fiber end. In this way, a simple compact MIR source can be realized and hence such sources can be a viable alternative to other types of MIR sources for some applications. So far, MIR fiber spontaneous emission sources have proved their ability for sensing greenhouse gases ($CO_2$, $CH_4$) [12-15].

In order to design efficient mid-infrared spontaneous emission sources a detailed spectroscopic analysis, which can enable numerical modeling and help better understanding of the behavior of MIR emission in $Pr^{3+}/Dy^{3+}$ co-doped selenide-chalcogenide fiber, is needed. The spectroscopic properties of co-doped $Dy^{3+}/Ho^{3+}$, $Dy^{3+}/Pr^{3+}$, $Dy^{3+}/Tb^{3+}$ and $Dy^{3+}/Tm^{3+}$ selenide-chalcogenide were studied by Park *et al.* [16]. The aim of this latter study was to enhance the 3 µm emission from $Dy^{3+}$ doped selenide-chalcogenide glasses in order to achieve efficient lasing at this wavelength. However, more recent show that there is a particularly large interest in the mid-infrared photoluminescence from $Pr^{3+}$ and $Dy^{3+}$ doped selenide-chalcogenide fibers. This is because co-doping $Pr^{3+}$ and $Dy^{3+}$ ions in a selenide-chalcogenide glass gives a high pump absorption cross-section in the near infrared spectral region [12-21]; these ions can be pumped with commercially available laser diodes operating at approximately 1.5 µm in the case of $Pr^{3+}$ and 1.3 µm and 1.7 µm in the case of $Dy^{3+}$. The ability to pump those doped glasses using commercially available laser diodes further enhances the prospect of practical application for MIR light generation. Lately, Ari *et. al.* presented *quasi*-constant emission in the 2.2-5.5 µm wavelength range from a co-doped $Pr^{3+}/Dy^{3+}$ sulfide fiber, under dual pumping at 0.915 µm and 1.55 µm [13]. The emission from this fiber was used for the simultaneous detection of $CO_2$ (4.26 µm) and $CH_4$ (3.34 µm). Due to the fact that selenide-chalcogenide glasses have lower phonon energy than sulfide glasses they should be better suited for the development of MIR spontaneous emission sources operating efficiently especially at longer wavelengths. Recently, 8 µm photoluminescence (PL) has been observed in $Tb^{3+}$ and $Sm^{3+}$ doped selenide-chalcogenide glass fibers [22, 23]. Thus, the main aim of the research presented in this contribution is to show that enhancement of the spontaneous emission can be achieved using a selenide-chalcogenide glass instead of sulfides. Furthermore, it is demonstrated that by varying the wavelength and output power of the pump lasers, the shape of the MIR spontaneous emission can be modified to either enhance or reduce the power content within a specified wavelength range.

In particular, in this contribution, the optical properties of in-house fabricated $Pr^{3+}/Dy^{3+}$ co-doped selenide-chalcogenide glass fiber were investigated experimentally. Pump lasers operating at 1.32 µm, 1.511 µm, 1.7 µm were used to excite the $Pr^{3+}/Dy^{3+}$ co-doped selenide-chalcogenide fiber. With a suitable arrangement of the pumping conditions, an output spectrum spanning from around 2 µm to 6 µm was recorded.

The paper is organized as follows: In Section 2, the experimental set-up is described; in Section 3, the photoluminescence (PL) emission spectra from $Pr^{3+}/Dy^{3+}$ co-doped fiber are reported under different pumping conditions, and the results obtained are discussed. Finally, conclusions are drawn in Section 4.

## 2. Experimental background

500 ppmw $Pr^{3+}$ and 500 ppmw $Dy^{3+}$ co-doped Ge-As-Ga-Se glass rods were prepared using the conventional melt-quenching method, more details of which can be found in [20, 21]. The glass composition was $Ge_{14.9}As_{20.9}Se_{62.7}Ga_{1.5}$ and more information about the glass composition development can be found in [24]. The preform was fiber-drawn into 320 μm diameter unstructured fiber and was not polymer-coated. More details about chalcogenide glass, rare earth ion doped, multimode fiber fabrication can be found in [20, 21].

A basic, schematic diagram of the experimental set-up for measuring the fluorescent decay and output power is presented in Fig. 1. For measuring PL spectra, a chopper and lock-in amplifier were added to the setup. PL spectra were obtained in the MIR spectral region by excitation of the $Pr^{3+}/Dy^{3+}$ fiber, using different pumping sources. In order to pump this fiber 1.32 μm (SemiNEX 4PN-116), 1.511 μm (SemiNex 4-PN-109) and 1.7 μm (QPC Laser PR-6017-0000) laser diodes were used. Dual wavelength pumping was achieved by combining two pump sources using a commercially available multimode fiber optic coupler (Thorlabs, TM200R2S2B). The collected PL was focused on to the slit of a motorized Lot Quantum 1/8 m monochromator, with a diffraction grating blazed at 4 μm, using a pair of lenses consisting of a BD (Black Diamond) lens with f=1.873 mm (Thorlabs C037ME-E) and $CaF_2$ with f=50 mm. The PL signal was modulated by using a chopper (Scitec Instruments), providing the reference signal for a lock-in amplifier. Detection of the signal was achieved with a thermoelectric cooled (200 K) MCT (mercury cadmium telluride) detector (Vigo System PVI-5). This MIR detector operates in the spectral range between 2 μm-6.5 μm. A lock-in amplifier (Scitec Instruments 410) was used to increase the signal-to-noise ratio. The measured spectra were recorded, and stored using a data acquisition card (NI USB-6008 National Instruments) and a computer. Emission spectra were recorded over the range of wavelengths: 2 μm-6 μm. All fluorescence spectra were measured at 300 K. In order to remove unwanted pump light, and also higher order diffraction grating contributions, long pass filters were used. In this case two long pass filters were used: a germanium window which acted as a long-pass filter with cut-on wavelength around 2 μm and an Edmund Optics long-pass filter with a cut-on wavelength of 3.5 μm. A Globar© blackbody source was used to correct for the system response for all the spectra presented in this contribution. The correction for the system response was preformed according to the procedure described in [3]. The PL decay rates were measured at the fiber end using the MCT detector and directly modulating the pump lasers. The decay rates were measured at the wavelengths: 2.4 μm, 2.95 μm, 4.4 μm and 4.7 μm. In the case of a singly-doped lanthanide ion doped glass host, with either $Pr^{3+}$ or $Dy^{3+}$ ions, the PL and decay rates were measured from the fiber-side. The time response of the detector and the preamplifier used in the PL decay measurements was less than 1 μs. In order to discriminate between the measured wavelengths a monochromator was used to act as a tunable bandpass filter. Time evolution of the fluorescence was recorded using a digital oscilloscope (PicoScope 5442A). The measured fluorescence decays were collected up to 512 times using the digital oscilloscope and results averaged in order to improve the signal-to-noise ratio.

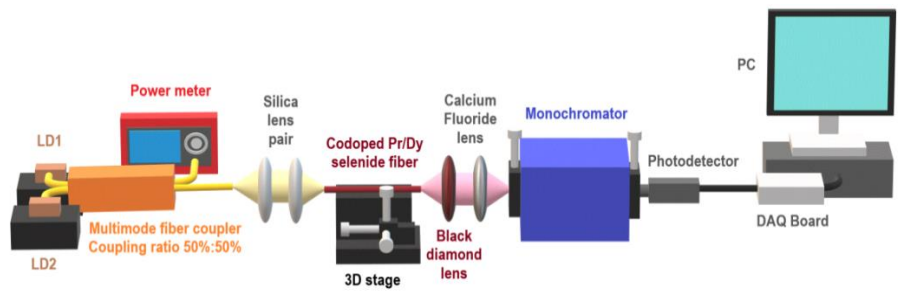

Fig. 1. Optical setup for measuring fiber photoluminescence (PL) properties under different pumping conditions.

## 3. Experimental results

The first two subsections of this section summarize the PL properties of singly doped $Pr^{3+}$ and $Dy^{3+}$ chalcogenide glass hosts. The following three subsections describe studies of the MIR emission spectra obtained from $Pr^{3+}/Dy^{3+}$ co-doped selenide-chalcogenide fiber under single wavelength pumping. The sixth subsection provides results for MIR spectra obtained from $Pr^{3+}/Dy^{3+}$ co-doped selenide-chalcogenide fiber samples using a combination of two pump wavelengths, whilst subsection seven contains a discussion of the measured photoluminescence lifetimes. Finally, subsection eight presents the measured dependence of MIR output power on differing pumping excitation.

Figure 2 shows a simplified energy level diagram of $Pr^{3+}$ and $Dy^{3+}$ ions doped into a selenide-chalcogenide glass host, up to an energy of 8000 cm$^{-1}$. The arrows indicate possible mid-infrared transitions that may occur when pumping using near-infrared lasers. In the case of $Pr^{3+}$ ions, when pumping at 1.511 µm, several transitions, e.g. $(^3F_4,^3F_3) \rightarrow (^3F_2,^3H_6)$, $(^3F_2,^3H_6) \rightarrow ^3H_5$ and $^3H_5 \rightarrow ^3H_4$, can contribute to the mid infrared emission at around 4.7 µm. Additionally, two transitions $(^3F_4,^3F_3) \rightarrow (^3H_5)$ and $(^3F_2,^3H_6) \rightarrow ^3H_4$ can contribute to emission at around 2.4 µm. Similarly, in the case of $Dy^{3+}$ when pumping at 1.32 µm, three MIR transitions centered at 5.4 µm $(^6F_{11/2}, ^6H_{9/2}) \rightarrow ^6H_{11/2}$, 4.4 µm, $^6H_{11/2} \rightarrow ^6H_{13/2}$ and 2.95 µm $^6H_{13/2} \rightarrow ^6H_{15/2}$ can occur. However, these transitions are well separated and can be easily distinguished, in contrast to $Pr^{3+}$. Also, emission at 2.4 µm can be expected from the $(^6F_{11/2}, ^6H_{9/2}) \rightarrow ^6H_{13/2}$ transition in $Dy^{3+}$.

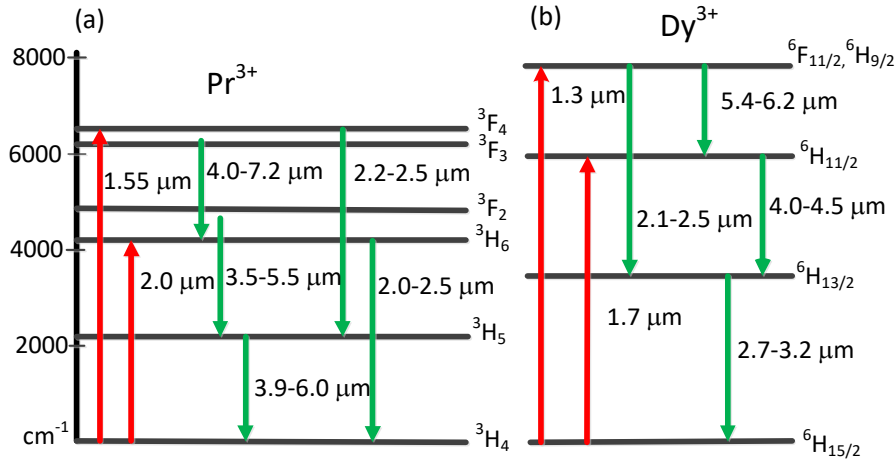

Fig. 2. Schematic energy level diagrams for: (a) $Pr^{3+}$ and (b) $Dy^{3+}$ doped, selenide-chalcogenide glass, up to an energy of 8000 cm$^{-1}$ (1.25 µm). The arrows indicate possible mid-infrared transitions.

### 3.1 MIR emission properties of Pr³⁺ chalcogenide glass

In this subsection, the PL properties of chalcogenide glass fiber doped with 1000 ppm $Pr^{3+}$ are summarized, whilst in subsection 3.2, 1000 ppm $Dy^{3+}$ doped chalcogenide glass fiber is considered. In both cases the fiber is unstructured and has a diameter of approximately 350 µm. This introductory discussion of the luminescent properties of single ion doped glasses helps better understanding of the behavior of MIR emission in $Pr^{3+}/Dy^{3+}$ co-doped selenide-chalcogenide fiber presented in subsections 3.3-3.8.

Figure 3(a-b) depicts the room temperature infrared emission bands of $Pr^{3+}$ GeAsSeGe glass in the wavelength ranges 3.25 µm-6 µm and 2 µm-3 µm, under excitation at 1.511 µm. PL spectra and lifetimes in this case were collected from the side of the fiber. The 3.5 µm-6 µm emission band is mainly associated with the $(^3F_2,^3H_6) \rightarrow ^3H_5$ and $^3H_5 \rightarrow ^3H_4$ transitions, however the emission from $(^3F_4,^3F_3) \rightarrow (^3F_2,^3H_6)$ can also contribute to the mid-infrared emission. Emission

recorded between 2 μm-3 μm can be associated with the overlapping levels $(^3F_4,^3F_3)\rightarrow{}^3H_5$ and $(^3F_2,^3H_6)\rightarrow{}^3H_4$ (*cf.* Fig 2a).

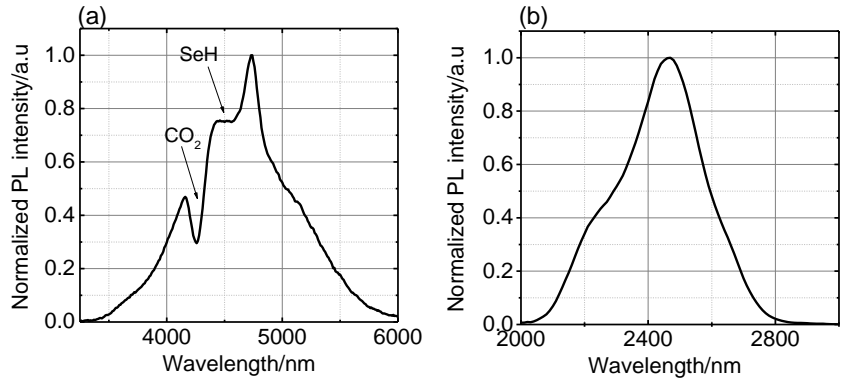

Fig. 3. Measured emission of $Pr^{3+}$ doped GeAsGaSe fiber: (a) from 3.5 to 6 μm (this emission can be attributed to the transitions $(^3F_2,^3H_6)\rightarrow{}^3H_5$, $^3H_5\rightarrow{}^3H_4$ and $(^3F_4,^3F_3)\rightarrow(^3F_2,^3H_6)$) and (b) from 2 to 3 μm (this emission can be attributed to two different transitions $(^3F_4,^3F_3\rightarrow{}^3H_5)$ and $(^3F_2,^3H_6\rightarrow{}^3H_4)$).

Figure 4(a) presents the PL decay characteristics of the 1000 ppmw $Pr^{3+}$ doped GeAsGaSe fiber, measured at 4.7 μm wavelength. In this case it was concluded that the best fit was obtained with a sum of two exponentials. The fast decay ($\tau_1$=1.93 ms) can be attributed to the $(^3F_2,^3H_6)\rightarrow{}^3H_5$ transition whilst the slow decay ($\tau_2$=7.6 ms) stems from a contribution of the $^3H_5\rightarrow{}^3H_4$ transition [20]. Figure 4(b) depicts the measured PL decay characteristic at 2.4 μm. In this case the best fit was also obtained with a sum of two exponentials, where the fast decay ($\tau_1$=0.255 ms) can be attributed to the $(^3F_4,^3F_3)\rightarrow(^3H_5)$ transition and the slow decay ($\tau_2$=2.2 ms) is a contribution from the $(^3F_2,^3H_6\rightarrow{}^3H_4)$ transition measured at 4.7 μm. Table 1 summarizes the measured PL decay lifetimes.

Measured in this work lifetimes of 7.6 ms for $^3H_5$, 2.065 ms for $(^3H_6, ^3F_2)$ and 0.255 ms for $(^3F_4, ^3F_3)$ levels compare well with those reported in literature [17,25-27], which are respectively: 8-12.5 ms for $^3H_5$ [17,25-27], 2.7 ms for $(^3H_6, ^3F_2)$ [17] and 0.25 ms for $(^3F_4, ^3F_3)$ [17, 27] levels. Moreover, the photoluminescence lifetimes measured in this contribution agree with our results of Judd-Ofelt (J-O) analysis performed for $Pr^{3+}$ and $Dy^{3+}$ doped selenide chalcogenide glasses which were presented in [28,29] and also with radiative lifetimes calculated by other authors using J-O analysis, which were: 9-15 ms for $^3H_5$ [17,28-31], 2.7-4.56 ms for $(^3H_6, ^3F_2)$ [17, 28-31] and 0.15-0.32 ms for $(^3F_4, ^3F_3)$ [17,27,30] levels, respectively.

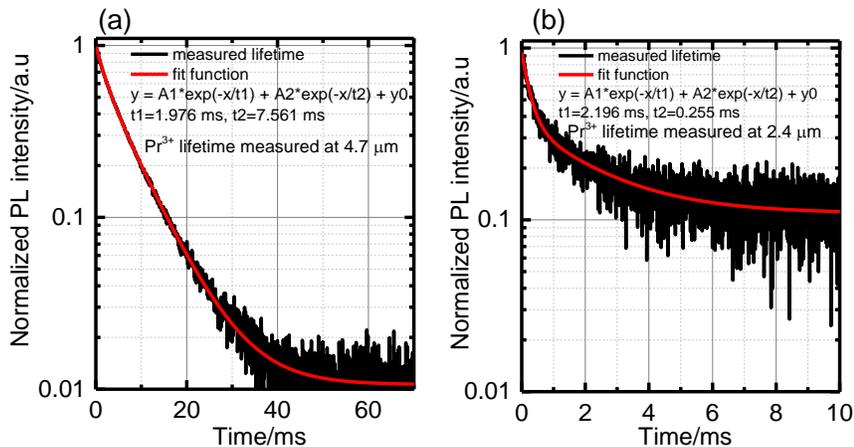

Fig. 4. Measured luminescence decay at: a) 4.7 μm and at b) 2.4 μm in 1000 ppm $Pr^{3+}$ GeAsGaSe chalcogenide glass fiber after the laser excitation at 1.511 μm.

Table 1. Measured PL lifetimes of $Pr^{3+}$: GeAsGaSe.

| Material | Upper State | Measured lifetime, τ / (ms) |
|---|---|---|
| $Pr^{3+}$: GeAsGaSe | $(^3F_4, ^3F_3)$ | 0.255 |
|  | $(^3F_2, ^3H_6)$ | 2.065 |
|  | $^3H_5$ | 7.6 |

## 3.2 MIR emission properties of $Dy^{3+}$ chalcogenide glass

Figure 5(a-b) displays the room temperature infrared emission bands of $Dy^{3+}$ doped GeAsSeGe glass in the wavelength ranges 3.5 μm-5 μm and 2 μm-3.5 μm, under excitation at 1.32 μm. The 3.5 μm-5 μm emission band is associated with the $^6H_{11/2} \rightarrow ^6H_{13/2}$ transition. PL spectra attributed to the $(^6F_{11/2}, ^6H_{9/2}) \rightarrow ^6H_{13/2}$ (2.1 μm–2.6 μm) and $^6H_{13/2} \rightarrow ^6H_{15/2}$ (2.7 μm – 3.4 μm) transitions are shown in Fig. 5b.

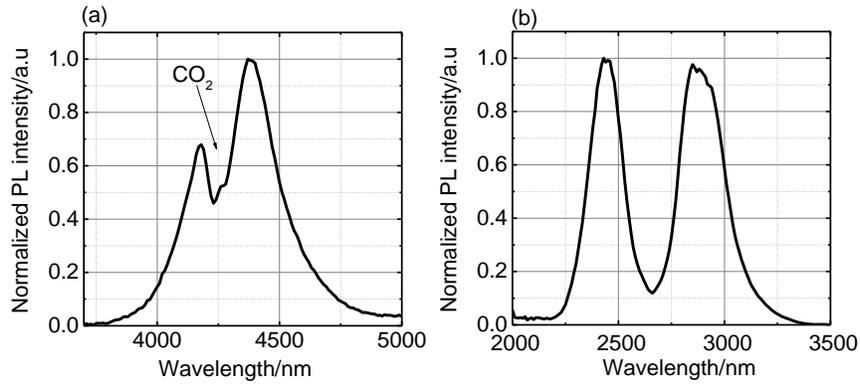

Fig. 5. Measured emission of $Dy^{3+}$ doped GeAsGaSe fiber: (a) from 3.5 to 6 μm (this emission can be attributed to the transition ($^6H_{11/2} \rightarrow ^6H_{13/2}$) and (b) from 2 to 3.5 μm (this emission can be attributed to two different transitions ($^6F_{11/2}, ^6H_{9/2}) \rightarrow ^6H_{13/2}$ (2.1–2.6 μm) and $^6H_{13/2} \rightarrow ^6H_{15/2}$ (2.7–3.4 μm)).

Figure 6(a-c) shows the photoluminescence decays for $Dy^{3+}$ doped GeAsGaSe fiber measured at the wavelengths of 2.95 μm, 4.4 μm and 2.4 μm, which correspond to the $^6H_{13/2} \rightarrow ^6H_{15/2}$, $^6H_{11/2} \rightarrow ^6H_{13/2}$ and $(^6F_{11/2}, ^6H_{9/2}) \rightarrow ^6H_{13/2}$ transitions, respectively. All the measured decays were best fitted with a single exponential. The measured PL lifetimes for $Dy^{3+}$ doped GeAsGaSe fiber are listed in Table 2.

In comparison with the literature data of $Dy^{3+}$-doped chalcogenide-selenide glasses the measured lifetimes of 3.8 ms for $^6H_{13/2}$, 1.74 ms for $^6H_{11/2}$ and 0.23 ms for $(^6F_{11/2}, ^6H_{9/2})$ levels obtained in this work are in good agreement with the measured lifetimes of 4.5-6.2 ms for $^6H_{13/2}$ [16,17], 2 ms for $^6H_{11/2}$ [17] and 0.31ms [17] for $(^6F_{11/2}, ^6H_{9/2})$ levels, respectively.

Moreover, the photoluminescence lifetimes measured in this contribution agree well with results of Judd-Ofelt (J-O) analysis, presented in [17, 28-31], which were: of 6.1-6.2 ms for $^6H_{13/2}$ [17,28], 2.2-2.4 ms for $^6H_{11/2}$ [17,28] and 0.38 ms [17] for $(^6F_{11/2}, ^6H_{9/2})$, respectively.

Based on comparison between measured and calculated lifetime it can be concluded that a quantum efficiency in all consider cases was higher than 50%, thus contribution of non-radiative process is low.

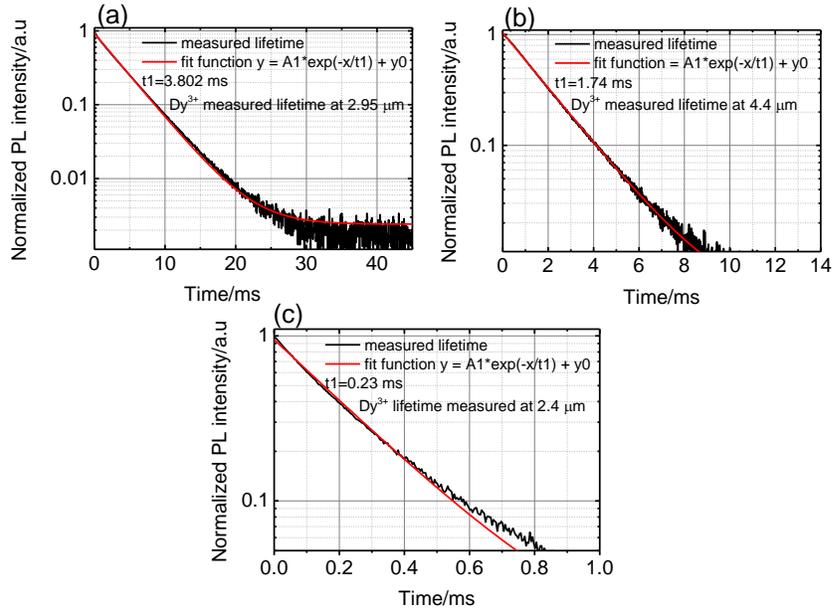

Fig. 6. Measured luminescence decay at: a) 2.95 µm and at b) 4.4 µm and c) 2.4 µm in 1000 ppm $Dy^{3+}$ GeAsGaSe chalcogenide glass fiber after the laser excitation at 1.32 µm.

Table 2. Measured PL lifetimes of $Dy^{3+}$: GeAsGaSe.

| Material | Upper State | Measured lifetime, $\tau$ / (ms) |
|---|---|---|
| $Dy^{3+}$:GeAsGaSe | ($^6F_{11/2}$, $^6H_{9/2}$) | 0.23 |
|  | $^6H_{11/2}$ | 1.74 |
|  | $^6H_{13/2}$ | 3.8 |

### 3.3 Pumping $Pr^{3+}/Dy^{3+}$ co-doped selenide-chalcogenide glass fiber at 1.32 µm

The measured emission from co-doped $Pr^{3+}/Dy^{3+}$ Ge-As-Ga-Se glass fiber under excitation at 1.32 µm is presented in Fig. 7. In order to facilitate the comparison of our results with the data available in the literature we present the recorded spectra both on linear and logarithmic scale [13]. The unclad fiber used in the experiment had a core diameter of *ca.* 320 µm and a length of 75 mm. The incident pump power was increased from 193 mW up to 530 mW. The PL was collected from the fiber end opposite to the pumped end. The broadband emission from around 2.1 µm up to 6 µm was recorded. The emission centered at the wavelength of 2.4 µm can be attributed to the ($^6F_{11/2}$, $^6H_{9/2}$)→$^6H_{13/2}$ transition from $Dy^{3+}$ and also due to two transitions ($^3F_4$,$^3F_3$)→($^3H_5$) and ($^3F_2$,$^3H_6$)→$^3H_4$ from $Pr^{3+}$. The PL at around 2.95 µm is attributable to emission from $^6H_{13/2}$→$^6H_{15/2}$ transition in $Dy^{3+}$. The emission centered at 4.4 µm should mainly consist of contribution from the $^6H_{11/2}$→$^6H_{13/2}$ transition in $Dy^{3+}$ and from the ($^3F_2$,$^3H_6$)→$^3H_5$, $^3H_5$→$^3H_4$ transitions in $Pr^{3+}$. However, under 1.32 µm pumping there is a 'gap' in the emission spectrum from around 3.4 µm to 3.7 µm. Additionally, the MIR emission spanning from 3.7 µm up to 6 µm is relatively weak in comparison with the emission centered at 2.95 µm. This behavior can be explained by the fact that, under 1.32 µm pumping, the emission centered at 4.4 µm is mainly due to the $^6H_{11/2}$→$^6H_{13/2}$ transition in $Dy^{3+}$ whilst the level $^6H_{11/2}$ is fed by the ($^6F_{11/2}$, $^6H_{9/2}$) transition, which has a low branching ratio of approximately 4 % to $^6H_{11/2}$ level. [17]. There is also a peak emission centered at 4.7 µm that is typical of $Pr^{3+}$ PL (*cf.* Fig.3a). Dips in the spectrum shown in Fig. 7 occurred at 2.7 µm and 4.26 µm, and can be attributed respectively to $H_2O$ vapor and $CO_2$ absorption in the ambient atmosphere. It should be also mentioned that the emission spectra of lanthanide ions doped into a chalcogenide selenide glass can be distorted by the presence of underlying OH and Se-H impurities present in the glass host, which have their respective absorption bands peaking at ~3 µm and ~ 4.5 µm (*e.g.* Fig.3a) [11].

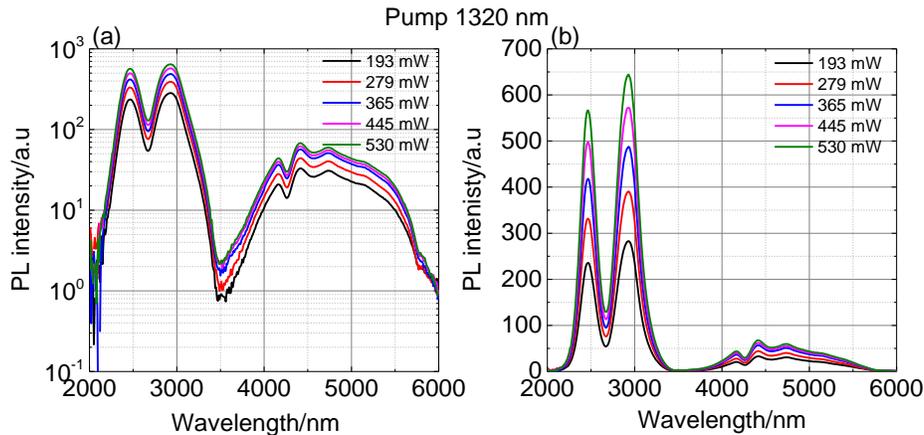

Fig. 7. (a) Measured infrared emission spectra on a logarithmic scale of co-doped 500 ppmw $Pr^{3+}$ and 500 ppmw $Dy^{3+}$ selenide-chalcogenide glass fiber under excitation at 1.32 µm recorded for different pump powers and (b) on a linear scale. The emission intensities were corrected for the system response.

### *3.4 Pumping $Pr^{3+}/Dy^{3+}$ co-doped selenide-chalcogenide glass fiber at 1.511 µm*

Figure 8 shows measured MIR emission spectra stretching from 2 µm to 6 µm when pumping using a 1.511 µm laser diode. The incident pump power illuminating the fiber end was varied between 85 mW and 285 mW. Under 1.511 µm pumping, emission might be expected to be mainly from the $Pr^{3+}$ ions, due to the fact that only $Pr^{3+}$ ions have an absorption band at 1.511 µm. However, in the results obtained experimentally an emission at 2.95 µm was also present (Fig. 8). This emission can be attributed to the $^6H_{13/2} \rightarrow {}^6H_{15/2}$ transition in the $Dy^{3+}$ (*cf.* Fig.5(b)). Therefore, this result directly confirms that an energy transfer between $Pr^{3+}$ and $Dy^{3+}$ ions has occurred. In comparison with the MIR emission measured under 1.32 µm pumping (Fig. 7), the MIR emission within the spectral region between 3.5 µm to 6 µm is considerably stronger. This can be explained by the fact that transitions with high branching ratio $(^3F_2, {}^3H_6) \rightarrow {}^3H_5$ (branching ratio around 44 %) and $^3H_5 \rightarrow {}^3H_4$ (branching ratio equal to 100 %) in $Pr^{3+}$ are excited directly by pumping at 1.511 µm [17].

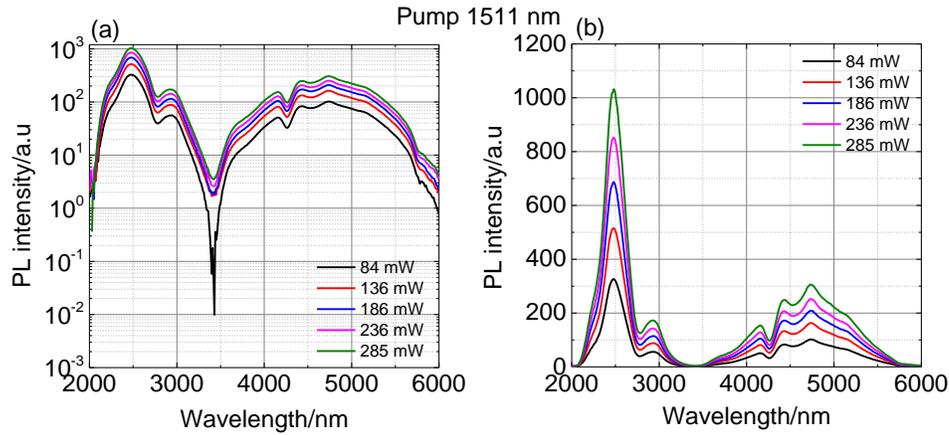

Fig. 8. (a) Measured infrared emission spectra on a logarithmic scale of co-doped 500 ppmw $Pr^{3+}$ and 500 ppmw $Dy^{3+}$ selenide-chalcogenide glass fiber under excitation at 1.511 μm recorded for different pump powers and (b) on a linear scale. The emission intensities were corrected for the system response.

### 3.5 Pumping $Pr^{3+}/Dy^{3+}$ co-doped selenide-chalcogenide glass fiber at 1.7 μm

Figure 9 shows the PL spectrum recorded for the wavelength range spanning from 2 μm to 6 μm under 1.7 μm pumping. The incident pump power was varied between 75 mW and 300 mW. It is noted that the absorption band $^3H_4 \rightarrow (^3F_4,^3F_3)$ centered at 1.6 μm in $Pr^{3+}$ and the absorption band $^6H_{15/2} \rightarrow ^6H_{11/2}$ centered at 1.7 μm in $Dy^{3+}$ overlap in the co-doped $Pr^{3+}/Dy^{3+}$ selenide chalcogenide glass [13]. Therefore, by pumping at 1.7 μm, $Pr^{3+}$ and $Dy^{3+}$ can be simultaneously directly excited. The results from Fig. 9 show that the emission centered around 2.95 μm makes a relatively larger contribution to the overall spectral content in comparison with the 1.511 μm pumping results presented in Fig. 8(a-b). Additionally, MIR emission in the range of 3.5 μm-6 μm is stronger than in case of 1.32 μm pumping. This can be explained by the fact that 1.7 μm pump light directly populates $^6H_{11/2}$ level in $Dy^{3+}$.

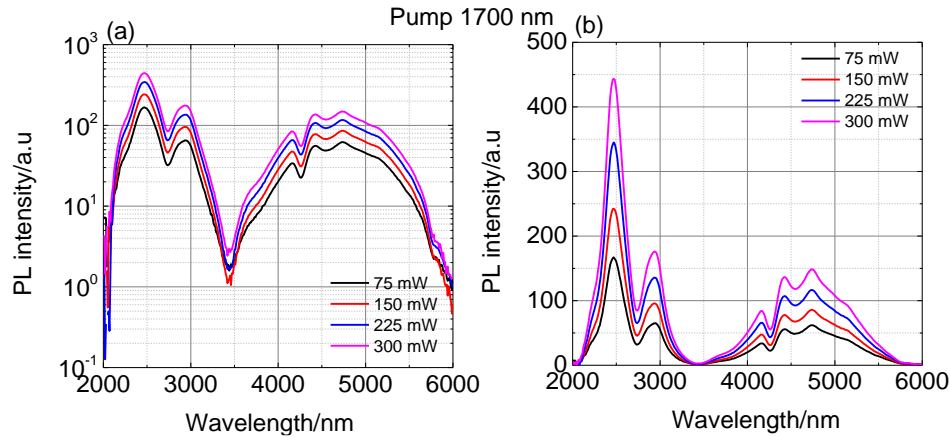

Fig. 9. (a) Measured infrared emission spectra of co-doped 500 ppmw $Pr^{3+}$ and 500 ppmw $Dy^{3+}$ selenide-chalcogenide glass fiber under excitation at 1.7 μm recorded for different pump powers on a logarithmic scale and (b) on a linear scale. The emission intensities were corrected for the system response.

### 3.6 Dual-pumping $Pr^{3+}/Dy^{3+}$ co-doped selenide-chalcogenide glass fiber

Ari *et al.* showed that the broadband mid-infrared spectral distribution in the case of a co-doped $Pr^{3+}/Dy^{3+}$ sulfide-chalcogenide fiber was obtained under dual-wavelength pumping at 0.915 μm and 1.55 μm wavelengths. Note that high energy electronic absorption edge of chalcogenide selenide glass is large at 0.915 µm wavelength. According to [32,33] fiber losses within this wavelength range are higher than 20 dB/m. Thus lasers operating at 0.915 µm are not suitable for pumping selenide-chalcogenide glass fibers studied in this paper. Therefore, the shortest pump wavelength considered in this study is 1.32 µm. Figure 10(a-b) presents the PL measured under the dual wavelength pumping (1.32 µm and 1.511 µm). Under dual-wavelength excitation the power ratio between the 1.32 µm laser diode and the 1.511 µm laser diode was kept at 1:1. In this case transitions at 2.4 µm and 2.95 µm contributed comparatively more to the total PL compared to the PL measured under 1.511 µm excitation. Additionally, the contribution of PL occurring between 3.5 µm-6 µm was also relatively larger when compared to the PL recorded under 1.32 µm excitation. Thus, by dual wavelength pumping, the main features of single-wavelength pumping at 1.32 µm and 1.511 µm can be combined. The minimum PL intensity occurs at around 3.42 µm. However, the PL intensity at 3.3 µm is low but still non-negligible. Thus, it can be used for $CH_4$ detection [13]. It is concluded that transitions at 2.4 µm and 2.95 µm contribute comparably under the dual wavelength pumping. Figure 11(a-d) shows the measured PL under dual wavelength pumping at 1.32 µm and 1.511 µm with different pump power ratios. These measurements were performed for two cases. In case (i), the pump power at 1.32 µm was held constant at the level of 75 mW while the pump power at 1.511 µm was varied from 150 mW up to 300 mW. In case (ii), the pump power at 1.511 µm was held constant at 75 mW while the pump power at 1.32 µm was varied from 150 mW up to 300 mW. From Fig. 11 (a-b) it can be noted that output power spectrum at 2.95 µm wavelength is almost constant whereas the intensity of the part of the spectrum spanning from 3.5-6 µm linearly increases with increasing the pump power at the fixed wavelength of 1.511 µm. The opposite situation takes place where the pump power at 1.511 µm is fixed at a constant power level. In this case the output power in the spectral range from 3.5 µm to 6 µm was almost constant whereas the output power at around 2.95 µm linearly increased with pump power at 1.32 µm (see Fig 11 (c-d). Thus, the results from Fig.11 demonstrate particularly that, by varying the output power and wavelength of the pump sources, the spectral shape of the emitted luminescence could be modified to either reduce or enhance the contribution of radiation within a specified wavelength band. It is interesting to note that in both cases the band centered at 2.4 µm was relatively strong. This can be explained by the fact that transitions centered at around 2.4 µm (($^6F_{11/2}$, $^6H_{9/2}$)→$^6H_{13/2}$ in $Dy^{3+}$ and ($^3F_4$,$^3F_3$→$^3H_5$) in $Pr^{3+}$ ) were pumped directly under excitation both at 1.32 µm and 1.511 µm.

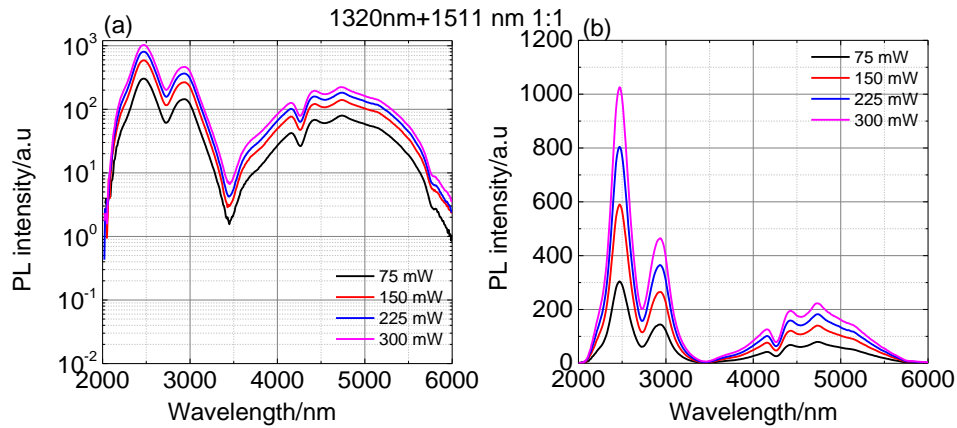

Fig. 10. (a) Measured infrared emission spectra of co-doped 500 ppmw $Pr^{3+}$ and 500 ppmw $Dy^{3+}$ selenide-chalcogenide glass fiber under dual-wavelength excitation at 1.32 µm and 1.511 µm on a logarithmic scale and (b) on a linear scale. The emission intensities were corrected for the system response.

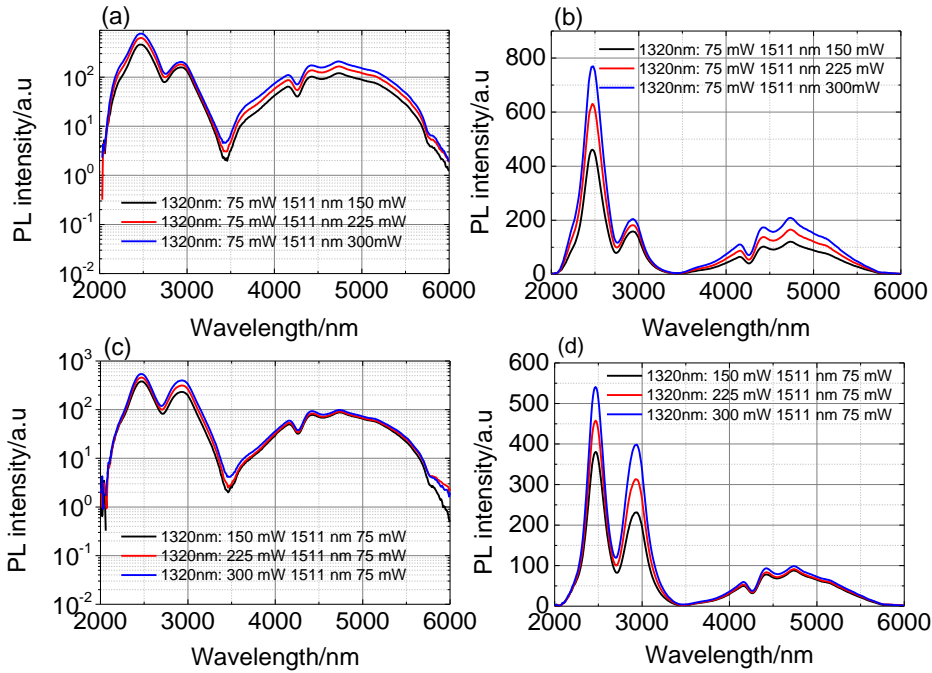

Fig. 11. Measured infrared emission spectra of co-doped 500 ppmw $Pr^{3+}$ and 500 ppmw $Dy^{3+}$ selenide-chalcogenide glass fiber under dual-wavelength excitation at 1.32 µm and 1.511 µm. (a) Pump power at 1.32 µm was constant at the level of 75 mW while the pump power at 1.511 µm was varied from 150 mW up to 300 mW plotted on a logarithmic scale and (b) plotted on a linear scale. Pump power at 1.511 µm was constant at the level of 75 mW while the pump power at 1.32 µm was varied from 150 mW up to 300 mW plotted on a logarithmic scale (c) and (d) plotted on a linear scale. The emission intensities were corrected for the system response

### 3.7 Fiber photoluminescence (PL) lifetimes

In order to understand better the possible transfer mechanisms in co-doped $Pr^{3+}$ and $Dy^{3+}$ selenide-chalcogenide glass PL, decays of various transitions were measured. The PL decay measurements show also that the photoluminescence lifetimes are large, thus demonstrating that the MIR radiative transitions were not suppressed by non-radiative transitions. Figure 12 presents the PL decay characteristics of $Pr^{3+}/Dy^{3+}$ co-doped Ge-As-Ga-Se fiber measured at 2.95 µm (see Fig. 12a), 4.4 µm (see Fig. 12b), 4.7 µm (see Fig. 12c) and 2.4 µm (see Fig. 12d) under 1.32 µm pumping. The PL lifetimes were measured when collecting the light from the fiber end. Under 1.32 µm pumping, the $Dy^{3+}$ ions were predominantly excited; however a transfer energy between $Dy^{3+}$ and $Pr^{3+}$ is clearly observed in the emission spectrum (see Fig. 7). The measured PL decay at 2.95 µm (see Fig. 12a) can be fitted by using a single exponential function with a lifetime of 3.7 ms. This lifetime can be attributed mainly to the $^6H_{13/2} \rightarrow {}^6H_{15/2}$ transition in $Dy^{3+}$. The measured lifetimes at 4.4 µm and 4.7 µm are the sum of two or more exponential functions. It should be noted here that the lifetimes measured at these wavelengths are far longer than those previously reported for chalcogenide-selenide glasses doped with $Dy^{3+}$ alone. (The measured lifetime at 4.4 µm for chalcogenide-selenide glass doped with $Dy^{3+}$ was equal to 1.74 ms, see Table 2 in section 3.2). This indicates that $Pr^{3+}$ ions were also excited and contribute to the PL within the spectral wavelength range 3.7 µm and 5.5 µm. The observed lifetimes within this spectral region for $Dy^{3+}/Pr^{3+}$ co-doped glass are in the range of a few milliseconds, which is a desirable feature for the construction of spontaneous emission fiber sources since this indicates that the contribution of the non-radiative transition is small. The PL decay measured at a wavelength of 2.4 µm (see Fig. 12d) also possessed a multi-exponential character (three terms were included), which can be attributed to $(^6F_{11/2}, {}^6H_{9/2}) \rightarrow {}^6H_{13/2}$ transition from $Dy^{3+}$ and also to transitions $(^3F_4, {}^3F_3) \rightarrow (^3H_5)$ and $(^3F_2, {}^3H_6) \rightarrow {}^3H_4$ from $Pr^{3+}$.

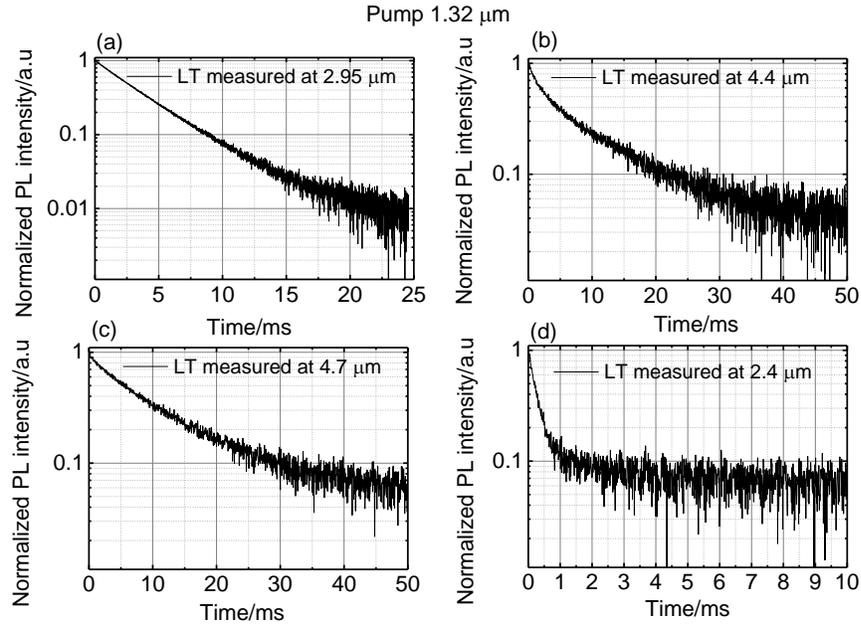

Fig. 12. PL decay of 500 ppmw $Pr^{3+}/Dy^{3+}$ co-doped GeAsGaSe fiber measured at: (a) 2.95 µm, (b) 4.4 µm, (c) 4.7 µm and (d) 2.4 µm under 1.32 µm pumping.

Fig. 13 shows the PL decays recorded at wavelengths: 2.95 µm, 4.4 µm, 4.7 µm and 2.4 µm under 1.511 µm pumping. Under 1.511 µm pumping mainly $Pr^{3+}$ ions should be excited. PL decay measured at 2.95 µm was fitted with a single exponential function of lifetime of 3.9 ms (similarly to the case of a $Dy^{3+}$ doped glass, *cf.* Table 2). This lifetime can be attributed to the $^6H_{13/2} \rightarrow {}^6H_{15/2}$ transition of $Dy^{3+}$. This confirms that with 1.511 µm pumping an energy transfer occurred between the $Pr^{3+}$ and $Dy^{3+}$ ions. As for 1.32 µm pumping, the observed infrared PL decay at 4.4 µm, 4.7 µm and 2.4 µm possessed a multi-exponential character with millisecond lifetimes, thus providing further confirmation that the contribution of non-radiative transitions was small.

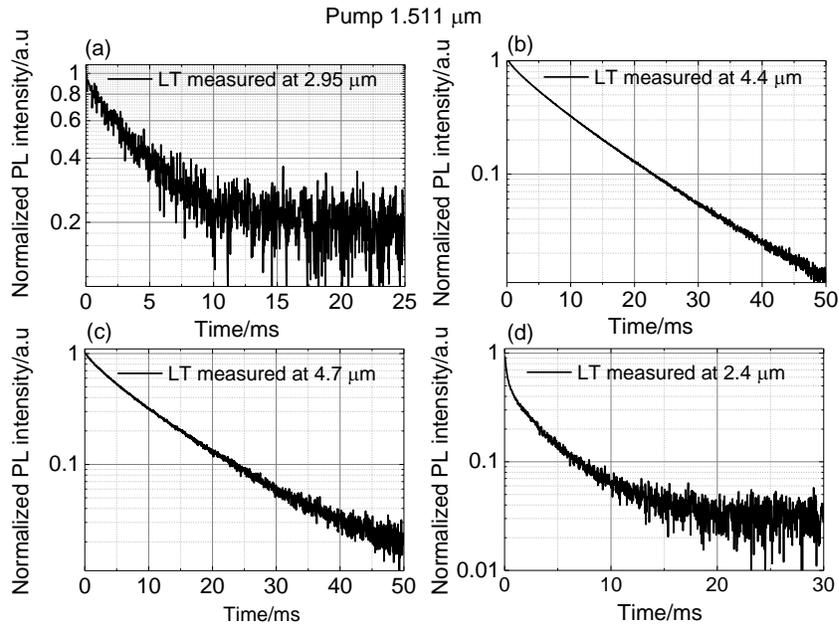

Fig. 13. PL decay of 500 ppmw Pr$^{3+}$/Dy$^{3+}$ co-doped Ge-As-Ga-Se fiber measured at: (a) 2.95 µm, (b) 4.4 µm, (c) 4.7 µm and (d) 2.4 µm all under 1.511 µm pumping.

*3.8 Output power measurements*

In order to characterize fully the potential of Pr$^{3+}$/Dy$^{3+}$ co-doped GeAsGaSe fiber for the realization of robust, low-cost spontaneous emission sources, the output power obtained under different pumping conditions was measured. The output power was measured by using a very low noise and high responsivity (up to 20 000 V/W) pyroelectric sensor (UM9B-BL-L-D0 Gentec.) The pump beam was modulated at a frequency of 10 Hz. The PL light was collected by a Black Diamond lens with NA=0.85 and f=1.873 mm (Thorlabs C037ME-E). The launching conditions were the same as presented in Fig.1, whilst the power meter photodetector replaced the calcium fluoride lens positioned in front of the monochromator (Fig.1). Two longwave pass filters with cut-on wavelengths 2 µm and 3.5 µm were used to eliminate the residual pump power. According to the power meter manufacturer specification the relative errors of power measurement are as high as 3%. Fig. 14(a) shows the dependence of the output MIR power on the 1.511 µm pump power. In this case the total output power in the wavelength range between 2 µm -6 µm was 191 µW whilst the output power above 3.5 µm was around 93 µW. This result is in a good agreement with measured spectral distribution under 1.511 µm pumping (see Fig. 8), where the calculated integral of power over the 3.5 µm-6 µm spectral range sums up to around 44% of the power integral calculated over 2 µm-6 µm band. A gentle sublinear behavior is observed in Fig.14(a). Similar saturation behavior was observed by other researchers performing mid-infrared study of rare earth ion doped chalcogenide glasses and was attributed to the phenomenon of ground state bleaching [14,34,35]. Figure 14(b) presents the dependence of the output MIR power on the 1.7 µm pump power. It is noted that the output fiber diameter of pumping laser diode at 1.7 µm was 400 µm whereas the Pr$^{3+}$/Dy$^{3+}$ co-doped GeAsGaSe fiber has a diameter 320 µm. Thus it can be estimated from geometric considerations than only 60% of pumping power at 1.7 µm is launched into the fiber. The maximum output power for the spectral region 2 µm-6 µm was 165 µW, with 67 µW of this stemming from spectral region above 3.5 µm.

Figure 14(c) shows the measured dependence of the output infrared power on the combined 1.511 µm and 1.32 µm pump power. In case of the dual pumping scheme the pump power ratio of 1.32 µm to 1.511 µm power was for these experiments kept at 1:1. The measured total output power was 135 µW for λ>2 µm and 51 µW for λ>3.5 µm.

The lowest output power was recorded under single excitation at 1.32 µm (see Fig.14(d)). The measured output power for λ>2 µm was only 88 µW of which 21 µW was at wavelengths λ>3.5 µm. To sum up, the results achieved, shown in Fig. 14(a)-(d), indicate that a MIR output power of 200 µW was obtained under excitation by commercial available laser diode operating in NIR spectral region.

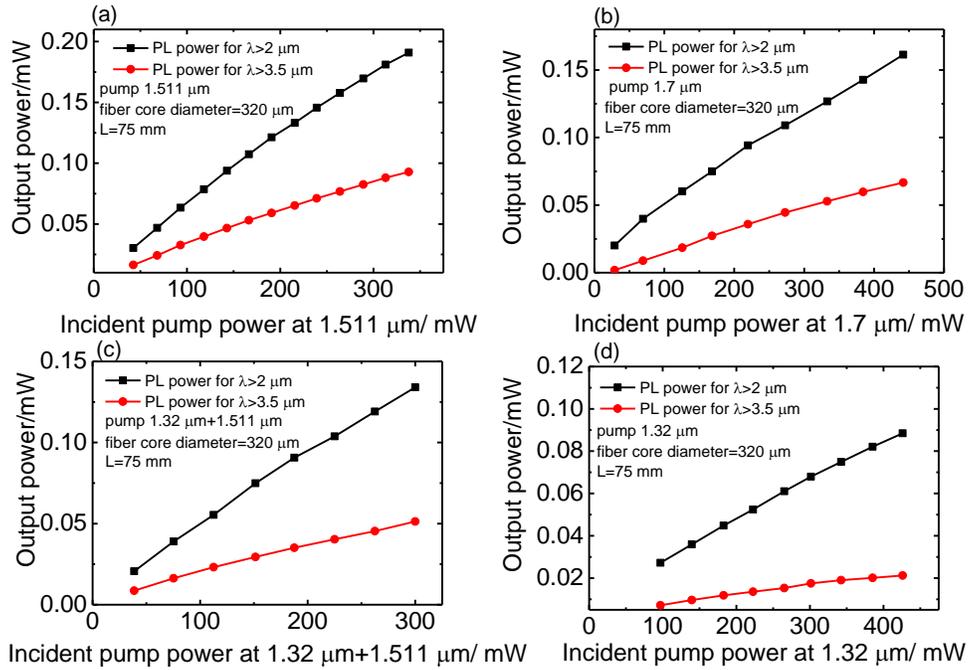
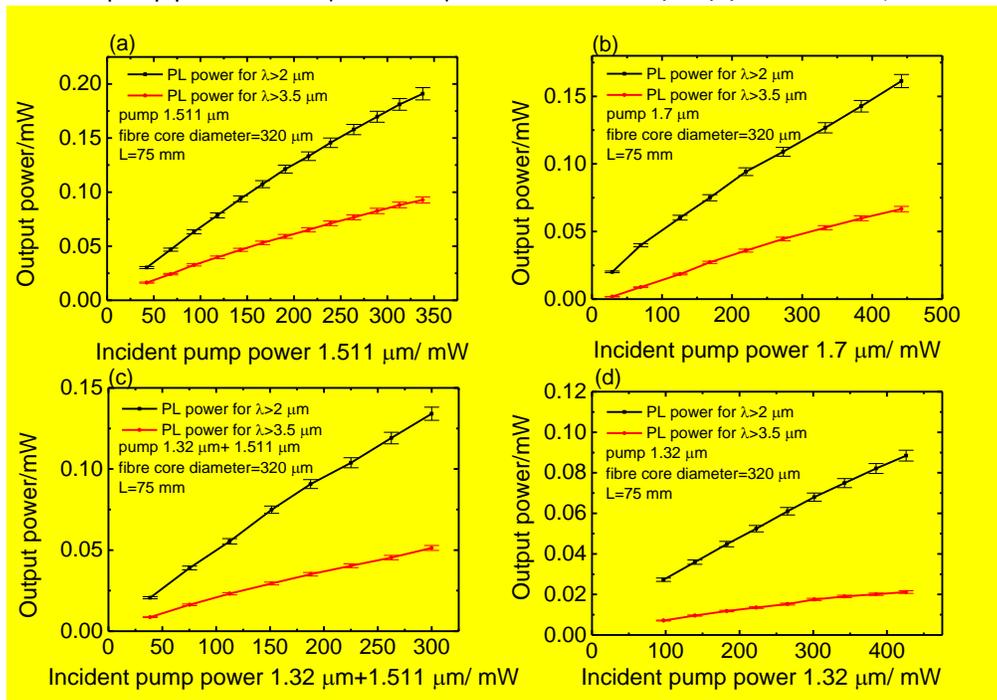

Fig. 14. Measured dependence of MIR output power on pump power for a pumping at (a) 1.511 µm, (b) 1.7 µm, (c) dual wavelength at 1.32 µm+1.511 µm and (d) 1.32 µm.

## 4. Conclusions

This work presented a spectroscopic study of 500 ppmw $Pr^{3+}/Dy^{3+}$ co-doped Ge-As-Ga-Se glass fibers pumped by commercially available laser diodes within the NIR part of the spectrum. A strong MIR emission in the range of 2000-6000 nm was observed. The PL emission properties of 500 ppmw $Pr^{3+}/Dy^{3+}$ co-doped Ge-As-Ga-Se glass fiber under different pumping wavelengths were investigated. Measured MIR lifetimes were in the range of a few milliseconds, confirming a low contribution from non-radiative processes. By applying various combinations of pump sources it was demonstrated that the MIR output spectrum can be shaped in order to either enhance or suppress the contribution of a particular band. Achieved results show the potential of $Pr^{3+}/Dy^{3+}$ co-doped Ge-As-Ga-Se glass fibers for the realization of low-cost robust broadband MIR spontaneous emission sources.

**Funding**

This project has received funding from a Newton Fund International Links Award (ref. 277109657) and the European Union's Horizon 2020 research and innovation programme under the Marie Skłodowska-Curie grant agreement No. 665778 (National Science Centre, Poland, Polonez Fellowship 2016/21/P/ST7/03666).